\documentclass[letterpaper, 10 pt, conference]{ieeeconf}  

\IEEEoverridecommandlockouts                              

\usepackage{graphicx}          
\usepackage{amsmath,amsfonts,amssymb}
\usepackage{amsmath}
\usepackage{comment}
\usepackage{threeparttable}

\newcommand{\red}[1]{\textcolor{black}{#1}}
\newcommand{\brown}[1]{\textcolor{black}{#1}}
\newcommand{\blue}[1]{\textcolor{black}{#1}}

\newcommand{\relx}[1]{\textcolor{black}{#1}}
\usepackage{makecell}
\usepackage{pbox}
\usepackage{subcaption}
\usepackage{xcolor}
\usepackage{tabularx}
\usepackage{cleveref}
\usepackage{float}

\crefname{equation}{Eq}{} 
\crefrangelabelformat{equation}{(#3#1#4--#5#2#6)}

\title{\LARGE \bf
\relx{Finite-volume method and observability analysis for core-shell enhanced single particle model for lithium iron phosphate batteries}
}

\author{Le Xu$^{1}$,  Simone Fasolato$^{2}$, and  Simona Onori$^{1,*}, \textit{IEEE Senior Member}$
\thanks{$^{1}$ Energy Science and Engineering, Stanford University, Palo Alto, USA.}
\thanks{$^{2}$ Department of Electrical, Computer and Biomedical Engineering, University of Pavia, Pavia, Italy.}
\thanks{$^{*}$ Corresponding author  {\tt\small sonori@stanford.edu}}}

\begin{document}

\maketitle
\thispagestyle{empty}
\pagestyle{empty}

\begin{abstract}
The increasing adoption of Lithium Iron Phosphate (LFP) batteries in Electric Vehicles is driven by their affordability, abundant material supply, and safety advantages. However, challenges arise in controlling/estimating unmeasurable LFP states such as state of charge (SOC), due to its flat open circuit voltage, hysteresis, and path dependence dynamics during intercalation and de-intercalation processes.
The Core-Shell Average Enhanced Single Particle Model (CSa-ESPM) effectively captures the electrochemical dynamics and phase transition behavior of LFP batteries by means of Partial Differential-Algebraic Equations (PDAEs). 
These governing PDAEs, including a moving boundary Ordinary Differential Equation (ODE), require a fine-grained spatial grid for accurate and stable solutions when employing the Finite Difference Method (FDM).
This, in turn, leads to a computationally expensive system intractable for the design of real-time battery management system algorithms.
In this study, we demonstrate that the Finite Volume Method (FVM) effectively discretizes the CSa-ESPM and provides accurate solutions with fewer than \blue{4} control volumes while ensuring mass conservation across multiple operational cycles. 
The resulting control-oriented reduced-order FVM-based CSa-ESPM is experimentally validated using various C-rate load profiles and its observability is assessed through nonlinear observability analysis. 
Our results reveal that different current inputs and discrete equation numbers influence model observability, with non-observable regions identified where solid-phase concentration gradients are negligible.
\end{abstract}

\section{Introduction}
Cobalt- and nickel-free positive electrode materials provide a pathway for building a resilient battery supply chain. Iron phosphate (${\text{LiFeP}}{{\text{O}}_{\text{4}}}$, also referred to as LFP) cathodes exhibit good thermal and chemical stability and also offer a better lifetime compared to NCA or NMC cathodes \cite{li2015modeling}. These characteristics make LFP batteries widely used in electrified transportation and gird applications. For LFP batteries, the presence of the two-phase transition results in a flat OCV curve which is also accompanied by a significant hysteresis during charge and discharge, making the modeling a challenging task. Among different electrochemical models, this study focuses on the core-shell average \blue{enhanced} single particle model (CSa-ESPM). The core-shell model is first proposed in \cite{Srinivasan_2004}, and the CSa-ESPM is proposed in \cite{Pozzato_2022,Pozzato_2022_csa}. After that, CSa-ESPM has been further investigated to reduce model complexity \cite{9991824} and account for hysteresis \cite{pozzato2024accelerating}.

The CSa-ESPM contains partial differential-algebraic equations (PDAEs) that need to be discretized and transformed into ordinary differential equations (ODEs) by using numerical methods. In previous studies, the finite difference method (FDM) is used to solve the CSa-ESPM with sufficient spatial grid resolution (i.e., 70 spatial discretization nodes). The resulting ODEs lead to a computationally expensive system for deploying on the battery management system (BMS). Using small spatial discretization nodes reduces the computational burden but the mass conservation property is not guaranteed when using FDM \cite{Xu_2023}.

In this study, the finite volume method (FVM), which conserves mass by design, is used to build a low-dimensional CSa-ESPM with less than 4 discretization control volumes, prone for on-board BMS applications. Model validation is conducted using experimental data at different C-rate, and non-linear observability analysis is studied for the positive electrode.
This paper is structured as follows. First, the CSa-ESPM model is introduced, and the FVM method is used to discretize and convert model equations into a system of ODEs. Then, experimental data is used to identify model parameters, followed by the nonlinear observability analysis of the CSa-ESPM model. Finally, conclusions are summarized in the last section.

\begin{figure}[!t]
	\centering
	\includegraphics[trim=0 0 10 0,clip,width = 1\columnwidth]{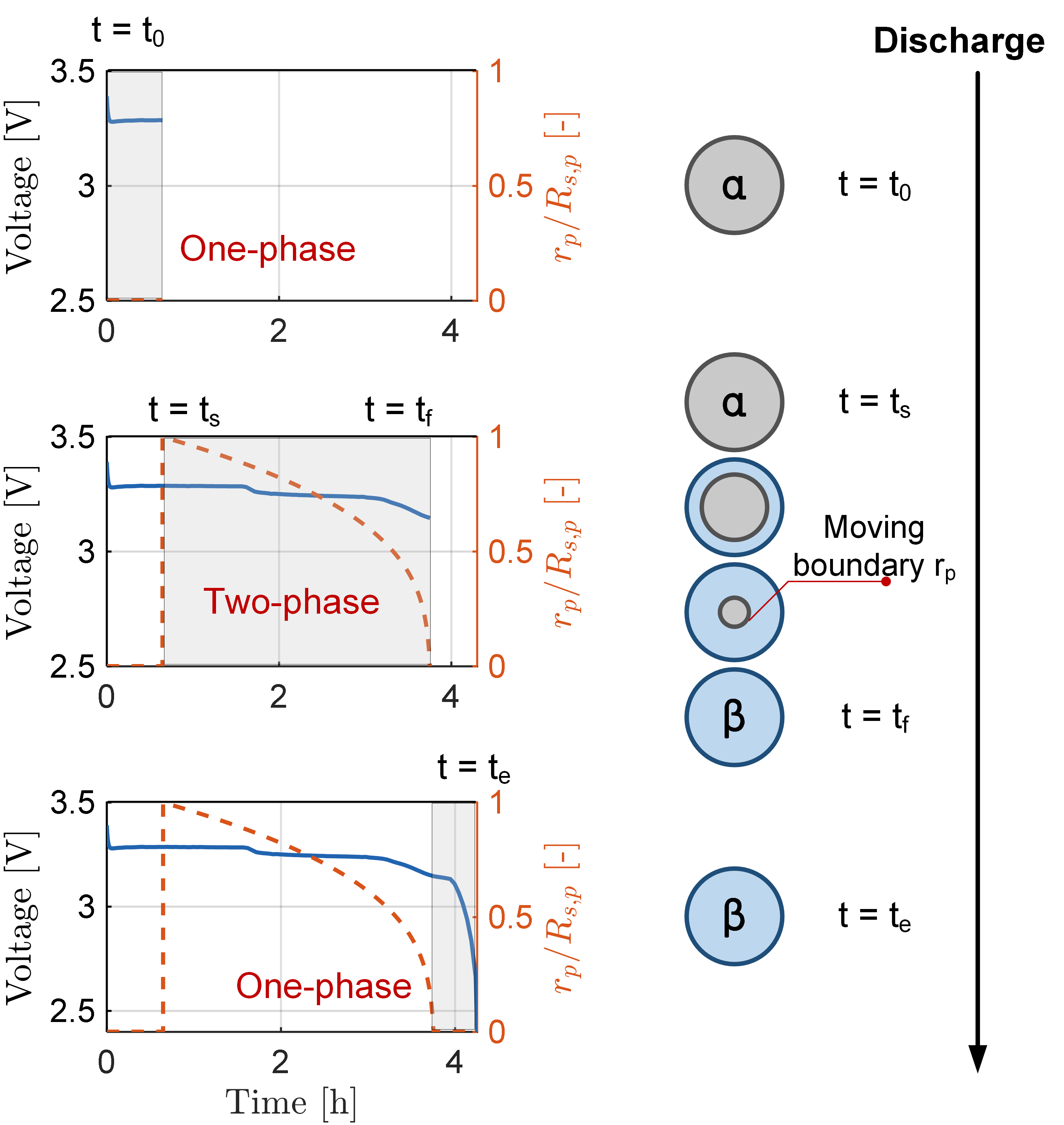}
	\caption{C/4 discharge. The moving boundary ($r_p$) is plotted to illustate the positive electrode one-phase ($r_p/R_{s,p}=0$) and two-phase ($r_p/R_{s,p}>0$) regions.}
	\label{fig:CoreShell_scheme}
	\vspace{-1.5 em}
\end{figure}

\begin{table*}
	\caption{CSa-ESPM discretized electrochemical governing dynamics}\label{Tab:POD}
	\centering
	    \begin{threeparttable}
	\resizebox{\textwidth}{!}{	
		\footnotesize{
			\begin{tabular}{ll}
				\hline \hline \\ [-1em]
				\multicolumn{2}{l}{Model state-space   formulation (The matrices and coefficients are reported in \cite{Pozzato_2022} and \cite{Xu_2023})} \\ \hline 
				Negative electrode and   one-phase positive electrode concentration   \cite{Xu_2023}         & Electrolyte concentration    \cite{Pozzato_2022}        \\
				\parbox{8.8cm}{ \begin{equation} \label{eq:tb1} {\mathbf{\dot{\bar{c}}}}_{s,j}^{} = {\mathbf{A}}_j^{}{\mathbf{\bar c}}_{s,j}^{} + {\mathbf{B}}_j^{}I, \quad \text{with:} \quad j = n,p \end{equation}}                                                                              
				& \parbox{8.8cm}{ \begin{equation} \label{eq:tb2} {\mathbf{\dot{\bar{c}}}}_e^{} = {\mathbf{A}}_e^{}{\mathbf{\bar c}}_e^{} + {\mathbf{B}}_e^{}I \end{equation}} \\ 
				\parbox{8.8cm}{ \begin{equation} \label{eq:tb1add} {\mathbf{\bar c}}_{s,j}^{} = {\left[ {{{\bar c}_{s1,j}},{{\bar c}_{s2,j}} \cdots {{\bar c}_{s{N_r},j}}} \right]^T} \in {\mathbb{R}^{{N_r} \times 1}} \text{with:} \quad j = n,p \end{equation}} &
				\parbox{8.8cm}{ \begin{equation} \label{eq:tb2add} {\mathbf{\bar c}}_{e}^{} = {\left[ {{{\bar c}_{e1}},{{\bar c}_{e2}} \cdots {{\bar c}_{e{N_e}}}} \right]^T} \in {\mathbb{R}^{{N_e} \times 1}}  \end{equation}}  \\
				\hline 
				\multicolumn{2}{l}{Positive electrode in   two-phase}                                                                 \\ 
				\hline
				Governing equations                                                            &                                      \\
				\parbox{8.8cm}{ \begin{equation} \label{eq:tb3} \frac{{\partial c_{s,p}^{}}}{{\partial t}} = \frac{{D_{s,p}^{}}}{{{r^2}}}\frac{\partial }{{\partial r}}\left( {{r^2}\frac{{\partial c_{s,p}^{}}}{{\partial r}}} \right) \end{equation}}                                                                              & \parbox{8.8cm}{ \begin{equation} \label{eq:tb4} \frac{{d{r_p}}}{{dt}} = \frac{{sign\left( I \right)D_{s,p}^{}}}{{\left( {c_{s,p}^\alpha  - c_{s,p}^\beta } \right)}}{\left. {\frac{{\partial {c_{s,p}}}}{{\partial r}}} \right|_{r = {r_p}}} \end{equation}}                                    \\
				State-space formulation                                                        &                                      \\
				\parbox{8.8cm}{ \begin{equation} \label{eq:tb5} {\mathbf{\dot{x}}}_{2,p}^{} = {\mathbf{A}}_s^{}{\mathbf{x}}_{2,p}^{} + {\mathbf{B}}_s^{}{I} + {{\mathbf{G}}_s} \end{equation}}        &   
				\parbox{8.8cm}{ \begin{equation} \label{eq:tb8} g\left( I \right) = \left\{ {\begin{array}{*{20}{c}}
								{\begin{array}{*{20}{c}}
										{c_{s,p}^\beta  = \theta _{p,\beta}  \cdot c_{s,p}^{\max },}&{\begin{array}{*{20}{c}}
												{{\text{if}}}&{I > 0} 
										\end{array}} 
								\end{array}} \\ 
								{\begin{array}{*{20}{c}}
										{c_{s,p}^\alpha  = \theta _{p,\alpha}  \cdot c_{s,p}^{\max },}&{\begin{array}{*{20}{c}}
												{{\text{if}}}&{I < 0} 
										\end{array}} 
								\end{array}} \\ 
								{\begin{array}{*{20}{c}}
										{0,}&{{\text{otherwise}}} 
								\end{array}} 
						\end{array}} \right. \end{equation}}                                                                             \\ 	\parbox{8.8cm}{ \begin{equation} \label{eq:tb6} {\mathbf{x}}_{2,p}^{} = {\left[ {{{\bar c}_{s1,p}},{{\bar c}_{s2,p}} \cdots {{\bar c}_{s{N_r},p}},r_p} \right]^T} \in {\mathbb{R}^{{(N_r+1)} \times 1}} \end{equation}}                                                               \\
				\multicolumn{2}{l}{	\parbox{18cm}{ \begin{equation} \label{eq:tb9} {A_s} = \frac{{3{D_{s,p}}}}{{\Delta {r_{{\text{2P}}}}}}{\left[ {\begin{array}{*{20}{c}}
										{ - \frac{{{{\left( {\Delta {r_{{\text{2P}}}} + {r_p}} \right)}^2} + 2r_p^2}}{{{{\left( {\Delta {r_{{\text{2P}}}} + {r_p}} \right)}^3} - r_p^3}}}& \cdots &0&0 \\ 
										\vdots & \ddots & \vdots & \vdots  \\ 
										0& \cdots &{\frac{{{{\left( {\left( {{N_r} - 1} \right)\Delta {r_{{\text{2P}}}} + {r_p}} \right)}^2}}}{{{{\left( {\left( {{N_r} - 1} \right)\Delta {r_{{\text{2P}}}} + {r_p}} \right)}^3} - R_{s,p}^3}}}&0 \\ 
										{\frac{{2\operatorname{sign} \left( I \right)}}{{3\left( {c_{s,p}^\alpha  - c_{s,p}^\beta } \right)}}} & \cdots &0&0
								\end{array}} \right]_{\left( {\left( {{N_r} + 1} \right) \times \left( {{N_r} + 1} \right)} \right)}} \end{equation}} }                                                                                                 \\
				\multicolumn{2}{l}{	\parbox{18cm}{ \begin{equation} \label{eq:tb10} {B_s} = \frac{3}{{{A_{cell}}F{L_p}{a_p}}}{\left[ {\begin{array}{*{20}{c}}
										0 \\ 
										\vdots  \\ 
										{\frac{{R_{s,p}^2}}{{{{\left( {\left( {{N_r} - 1} \right)\Delta {r_{{\text{2P}}}} + {r_p}} \right)}^3} - R_{s,p}^3}}} \\ 
										0 
								\end{array}} \right]_{\left( {\left( {{N_r} + 1} \right) \times 1} \right)}},{G_s} = {\left[ {\begin{array}{*{20}{c}}
										{ - \frac{{6{D_{s,p}}g\left( I \right)r_p^2}}{{\Delta {r_{{\text{2P}}}}\left( {{{\left( {\Delta {r_{{\text{2P}}}} + {r_p}} \right)}^3} - r_p^3} \right)}}} \\ 
										\vdots  \\ 
										0 \\ 
										{ - \frac{{2\operatorname{sign} \left( I \right){D_{s,p}}{\text{g}}\left( I \right)}}{{\Delta {r_{{\text{2P}}}}\left( {c_{s,p}^\alpha  - c_{s,p}^\beta } \right)}}} 
								\end{array}} \right]_{\left( {\left( {{N_r} + 1} \right) \times 1} \right)}} \end{equation}} }                                                                                                 \\ \hline
				Model output                                                                  &                                      \\ \hline
				Electrode overpotential                                                        & Electrolyte overpotential            \\
				\parbox{8.8cm}{ \begin{equation} \label{eq:tb11} \begin{gathered}
							{\eta _j} = \frac{{2RT}}{F}{\sinh ^{ - 1}}\left( {\frac{{Ip(I)}}{{2{a_{s,j}}{A_{cell}}{L_j}{i_{0,j}}}}} \right),j = n,p \hfill \\
							\begin{array}{*{20}{c}}
								{p(I) = \left\{ {\begin{array}{*{20}{l}}
											{ - 1,}&{j = p} \\ 
											{1,}&{j = n} 
									\end{array}} \right.,}&{{a_{s,j}} = \frac{3}{{{R_{s,j}}}}} 
							\end{array}{\varepsilon _j} \hfill \\ 
						\end{gathered}  \end{equation}}                                                                              & \parbox{8.8cm}{ \begin{equation} \label{eq:tb12} \Delta {\phi _e} = \frac{{2RTv}}{F}\ln \left( {\frac{{{c_e}\left( {{L_n} + {L_s} + {L_p}} \right)}}{{{c_e}\left( 0 \right)}}} \right) \end{equation}}                                    \\
				Exchange current density\\
				\parbox{8.8cm}{ \begin{equation} \label{eq:tbcom1} \begin{array}{*{20}{c}}
							{{i_{0,p}} = {k_p}F\sqrt {c_{e,p}^{avg}{c_{2p}}\left( {c_{s,p}^{max} - {c_{2p}}} \right)} }&{\text{with}\left\{ {\begin{array}{*{20}{c}}
										{{\text{One - phase}}:{c_{2p}} = c_{s,p}^{surf}} \\ 
										{{\text{Two - phase}}:{c_{2p}} = c_{s,p}^{bulk}} 
								\end{array}} \right.} 
						\end{array} \end{equation}} & \parbox{8.8cm}{ \begin{equation} \label{eq:tbcom2} {i_{0,n}} = {k_n}F\sqrt {c_{e,n}^{avg}c_{s,n}^{surf}\left( {c_{s,n}^{max} - c_{s,n}^{surf}} \right)}  \end{equation}} \\
				Cell voltage                                                                   & State of charge                      \\
				\parbox{8.8cm}{ \begin{equation} \label{eq:tb13} {V_{cell}} = {U_p} + {\eta _p} - {U_n} - {\eta _n} + \Delta {\phi _e} - {R_l}I \end{equation}}                                                                             & \parbox{8.8cm}{ \begin{equation} \label{eq:tb14} \begin{array}{*{20}{c}}
							{SOC_p^{ch/dis} = \frac{{\theta _{p,0}^{ch/dis} - \theta _p^{bulk}}}{{\theta _{p,0}^{ch/dis} - \theta _{p,100}^{ch/dis}}},}&{SOC_n^{ch/dis} = \frac{{\theta _n^{bulk} - \theta _{n,0}^{ch/dis}}}{{\theta _{n,100}^{ch/dis} - \theta _{n,0}^{ch/dis}}}} 
				\end{array} \end{equation}}                                    \\ \hline \hline
	\end{tabular}}} 
	\vspace{0em}
	       \begin{tablenotes}
		\item[1] $c_{s,j}^{surf}$ is the surface solid-phase concentration, and is given by Eq.42 and 43 in ref \cite{Xu_2023}.
		\item[2] $c_{s,j}^{bulk}$ and $\theta _{s,j}^{bulk}$ are the bulk and bulk-normalized solid-phase concentration, respectively. They are calculated using Eq.3 and 5 in ref \cite{Pozzato_2022_csa}.
	\end{tablenotes}
\end{threeparttable}
\end{table*}

\section{Core-shell average ESPM}\label{sec:Core_shell_ESPM}

In this section, a brief overview of the governing equations of the CSa-ESPM, emphasizing the phase transition mechanisms inherent in LFP batteries, is described. Then, the numerical solution of the positive electrode concentration diffusion governing equation, coupled with the moving boundary ODE, is introduced based on the FVM scheme. 
During operation, the positive electrode of LFP batteries experiences the formation of two phases \cite{willchue}: 1) a Li-poor phase ${\text{FeP}}{{\text{O}}_{\text{4}}}$, denoted as $\alpha$-phase, and 2) a Li-rich phase ${\text{LiFeP}}{{\text{O}}_{\text{4}}}$, denoted as $\beta$-phase. 
To facilitate the reader’s understanding of the phase transition mechanism within the CSa-ESPM, a schematic representation is provided in Fig. \ref{fig:CoreShell_scheme}.
At the initial time instant $t_0$, the LFP battery is fully charged and the positive electrode is in the $\alpha$-phase (Li-poor). As discharge continues, the concentration increases and reaches $c_{s,p}^\alpha  = \theta _{p,\alpha} \cdot c_{s,p}^{\max }$ at timestep $t_s$, then the formation of the $\beta$-phase (Li-rich) starts. In this stage, the coexistence of $\alpha$-phase and $\beta$-phase leads to a constant positive electrode open circuit potential (OCP). The transition from $\alpha$-phase to $\beta$-phase ends when concentration reaches $c_{s,p}^\beta  = \theta _{p,\beta} \cdot c_{s,p}^{\max }$ at timestep $t_f$. After this point, the positive electrode is in $\beta$-phase and the positive OCP decreases significantly until the end of discharge $t_e$.


\begin{figure*}[!h]
	\centering
	\includegraphics[trim=30 0 70 0,clip,width = 0.9\textwidth]{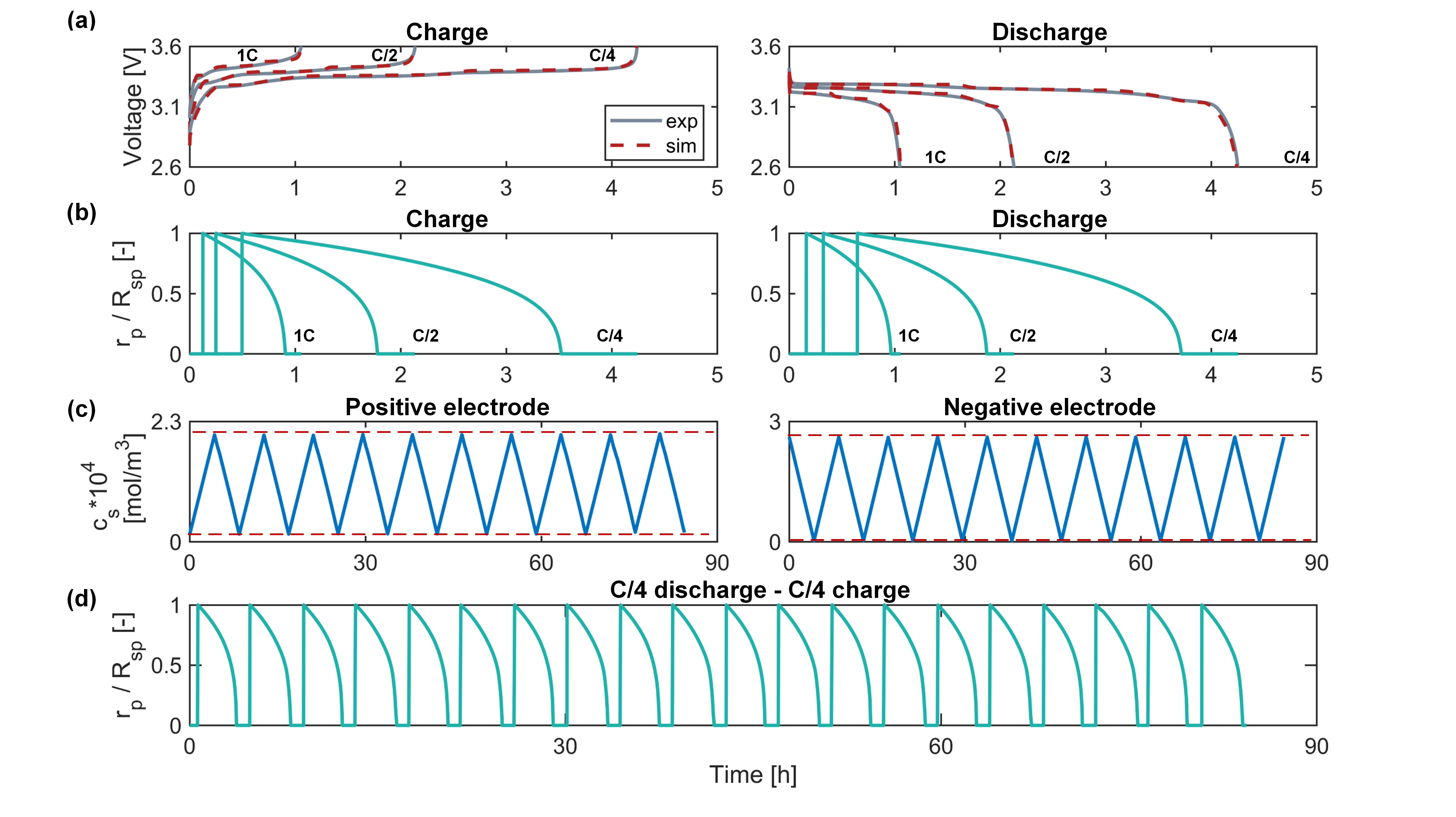}
	\caption{Comparison results at C/4, C/2, and 1C for (a) voltage, (b) moving boundary, (c) volume-average concentraion, (d) moving boundary during cycling.}
	\label{fig:iden_re}
	\vspace{0em}
\end{figure*}

\subsection{CSa-ESPM: Governing equations}
In the CSa-ESPM, the positive electrode in the two-phase region is modeled as one single particle with the core (Phase $\#$1) and the shell (Phase $\#$2). During discharge, for example, the concentration in the core is uniform and at a constant value $c_{s,p}^\alpha$. This is a shrinking process and the $\beta$-phase replaces the $\alpha$-phase as discharge takes place. The mass balance equation Eq.\eqref{eq:tb4} models this process where the motion of the $\alpha$- and $\beta$-phase interface, $r_p$, is assumed to be a function of the concentration gradient solely. Besides the positive particle diffusion equation, the remaining governing equations are the same as the ESPM. Table \ref{Tab:POD} summarizes the governing equations for CSa-ESPM, \red{where the subscripts $n$ and $p$ refer to the negative and positive electrodes, respectively.} The detailed model description for CSa-ESPM, including the electrode OCPs, is provided in \cite{Pozzato_2022}. In previous studies, the following coordinate transformation is proposed to remap the shell region from [$r_p$, $R_p$] to [0, 1], which makes the shell calculation domain stationary while the boundary is moving.

\begin{equation}
	\label{eq:FVM1} 
	\chi  = \frac{{r - {r_p}}}{{{R_p} - {r_p}}} \in \left[ {0,1} \right]
\end{equation}

After this transformation, the governing equations for the moving boundary and solid-phase diffusion in two-phase region becomes:
\begin{equation}
	\label{eq:FVM2} 
	\frac{{d{r_p}}}{{dt}} = \frac{{{\text{sign}}\left( I \right){D_{s,p}}}}{{\left( {c_{s,p}^\alpha  - c_{s,p}^\beta } \right)\left( {{R_p} - {r_p}} \right)}}{\left. {\frac{{\partial {c_{s,p}}}}{{\partial \chi }}} \right|_{\chi  = 0}}
\end{equation}
\begin{equation}
	\label{eq:FVM3} 
	\begin{gathered}
		\frac{{\partial {c_{s,p}}}}{{\partial t}} = \frac{{{\partial ^2}{c_{s,p}}}}{{\partial {\chi ^2}}}\left[ {\frac{{{D_{s,p}}}}{{{{\left( {{R_p} - {r_p}} \right)}^2}}}} \right] + \frac{{\partial {c_{s,p}}}}{{\partial \chi }}\left[ {\frac{{2{D_{s,p}}}}{{r\left( {{R_p} - {r_p}} \right)}}} \right] \hfill \\
		- \frac{{\partial {c_{s,p}}}}{{\partial \chi }}\frac{{\partial {r_p}}}{{\partial t}}\left[ {\frac{{\chi  - 1}}{{{R_p} - {r_p}}}} \right] \hfill \\ 
	\end{gathered}
\end{equation}

The complexity of the governing equations after coordinate transformation increases given that FVM requires the governing equation to be solved in its integral form. Next, we will illustrate how to solve the CSa-ESPM by FVM without coordinate transformation.

\subsection{CSa-ESPM: FVM scheme} \label{Sec:FVM_scheme}
In FVM scheme, the calculation domain is divided into control volumes (CVs), and the volume-averaged value of the solid-phase concentration (i.e., ${\bar c_{si,j}}$ \blue{with $,j = n,p$}) is calculated in each CV:
\begin{equation}
	\label{FVM1}
	{\bar c_{si,j}} = \frac{1}{{{V_i}}}\int_{C{V_i}} {{c_{si,j}}dV}
\end{equation}
where $V_i$ represents the volume of the $i^{th}$ CV.

The solid-phase diffusion equation in the integral form is solved using FVM scheme, thus guaranteeing mass conservation. By applying the Gauss theorem in spherical coordinates and using the second order difference scheme to approximate the diffusion terms (i.e., ${D_{s,j}}\left( {\frac{{\partial {c_{s,j}}}}{{\partial r}}} \right)$), the solid-phase diffusion equation for each $CV_i$ can be written as:
\begin{equation}
	\label{FVM2}
\begin{gathered}
	\frac{{\partial {{\bar c}_{si,j}}}}{{\partial t}}{V_i} = {D_{s,j}}\frac{{{{\bar c}_{s\left( {i + 1} \right),j}} - {{\bar c}_{si,j}}}}{{\Delta r}}{A_{i + \frac{1}{2}}} -  \hfill \\
	{D_{s,j}}\frac{{{{\bar c}_{si,j}} - {{\bar c}_{s\left( {i - 1} \right),j}}}}{{\Delta r}}{A_{i - \frac{1}{2}}},j = n,p \hfill \\ 
\end{gathered}
\end{equation}
Where index $i + \frac{1}{2}$ represents the interface between the $i^{th}$ and $i+1^{th}$ CV. ${A_{i - \frac{1}{2}}}$ and ${A_{i + \frac{1}{2}}}$ are the left and right surface areas of the $i^{th}$ CV, respectively.

In this study, $N_r$ refers to either the total number of CVs in FVM, or the spatial discretization nodes in FDM.

The state space representation for one-phase positive electrode and negative electrode concentration (Eq.\eqref{eq:tb1}) can be found in Eq.(39) - (41) in \cite{Xu_2023}.
In the two-phase region, the thickness of the positive electrode shell region is changing. Therefore, the length of each two-phase CV is a function of the moving boundary $r_p$ and is calculated as:
\begin{equation}
	\label{FVM3}
	\Delta {r_{{\text{2P}}}} = \frac{{{R_{s,p}} - {r_p}}}{{{N_r}}}
\end{equation}

Following the same approach outlined in \cite{Xu_2023}, the state space representation of Eq.\eqref{eq:tb3} and Eq.\eqref{eq:tb4} are given by Eq.\eqref{eq:tb5} - Eq.\eqref{eq:tb10}.

\section{Parameters identification} \label{Sec:PODGalerkin_MOR}
\red{Following the identification strategy proposed in} \cite{Pozzato_2022}, CSa-ESPM model parameters are identified. These parameters are shown in Table \ref{Tab:POD}. First, C/4 charge and discharge data are employed to identify the parameter vector, denoted as ${\lambda _{C/4}}$, which comprises:
\begin{equation}
	\label{pso1}
{\lambda _{C/4}} = \begin{array}{*{20}{c}}
	{\left[ {\theta _{p,100}^{ch}} \right.}&{\theta _{p,0}^{ch}}&{\theta _{n,100}^{ch}}&{\theta _{n,0}^{ch}}&{\theta _{p,\alpha }^{ch}}&{\theta _{p,\beta }^{ch}} \\ 
	{\theta _{p,100}^{dis}}&{\theta _{p,0}^{dis}}&{\theta _{n,100}^{dis}}&{\theta _{n,0}^{dis}}&{\theta _{p,\alpha }^{dis}}&{\theta _{p,\beta }^{dis}} \\ 
	{{R_{s,p}}}&{{R_{s,n}}}&{{D_{s,p}}}&{{D_{s,n}}}&{{\varepsilon _p}}&{{\varepsilon _n}} \\ 
	{{k_p}}&{{k_n}}&{{A_{cell}}}&{\left. {{R_l}} \right]}&{}&{} 
\end{array}
\end{equation}

It is important to highlight that the stoichiometric values differ between charge and discharge conditions. This discrepancy arises from the utilization of distinct positive OCP curves for charge and discharge simulations, as outlined in Equation (8) of \cite{9991824}. Consequently, separate calibration of the stoichiometric window is necessary.
Then, the following parameter vector is identified using C/2 and 1C charge data.

\begin{equation}
	\label{pso2}
	{\lambda _{C/2}} = {\lambda _{1C}} = \left[ {\begin{array}{*{20}{c}}
			{{D_{s,p}},}&{{D_{s,n}},}&{{k_p},}&{{k_n}} 
	\end{array}} \right]
\end{equation}

The identified values are shown in Table \ref{Tab:para_values}, and the identification results are shown in Fig. \ref{fig:iden_re}(a)-(c). Here, only 4 CVs are used to discretize the CSa-ESPM (FVM with $N_r=4$). 

Model validation is conducted using C/2 discharge and 1C discharge data. As can be seen from Fig. \ref{fig:iden_re}(a), the simulated voltage matches well with the measured data. Also, the simulated  moving boundary $r_p$ is shown in \ref{fig:iden_re}(b). The root-mean-square-error (RMSE) of voltage under C/2 and 1C discharge are 14.96 mV and 23.89 mV, respectively. Above results show that FVM-based CSa-ESPM has high accuracy with only 4 CVs. To the best of our knowledge, it is the first time that CSa-ESPM solved by FVM is validated using experimental data. 

Besides accuracy, the mass conservation property is also checked. Following the approach presented in \cite{Xu_2023}, we used a C/4 charge – C/4 discharge profile for multiple cycles simulation, which ensures the total ampere-hour throughput are the same for charge and discharge. Fig. \ref{fig:iden_re}(c) shows that the peak values for positive and negative electrodes volume-average concentration remain constant, which proves the mass is conserved when using FVM for solving CSa-ESPM. Also, Fig. \ref{fig:iden_re}(d) shows that the moving boundary changes continuously between one-phase and two-phase regions.  

	\begin{table}[!h]
	\centering
	\caption{Identified parameters at different C-rates.}	
	\label{Tab:para_values}
	\scriptsize{
		\begin{tabular}{ccccc}
			\hline \hline
			Current profiles & \multicolumn{3}{c}{Charge}         & Discharge \\ \hline
			C-rates          & C/4 & C/2 & 1C                     & C/4       \\ \hline
			${\theta _{n,100\% }}$ [-]          & 0.832   & -   & \multicolumn{1}{c|}{-} & 0.831  \\
			${\theta _{n,0\% }}$   [-]        & 0.011   & -   & \multicolumn{1}{c|}{-} & 0.009   \\
			${\theta _{p,100\% }}$   [-]        & 0.065   & -   & \multicolumn{1}{c|}{-} & 0.066   \\
			${\theta _{p,0\% }}$    [-]       & 0.910   & -   & \multicolumn{1}{c|}{-} & 0.925  \\
			${\theta _{p,\alpha}}$  [-]       & 0.220   & -   & \multicolumn{1}{c|}{-} & 0.196   \\
			${\theta _{p,\beta}}$   [-]       & 0.817   & -   & \multicolumn{1}{c|}{-} & 0.804   \\
			${R_{s,n}}$   [m]     & 8.10e-07   & -   & \multicolumn{1}{c|}{-} & 8.10e-07    \\
			${R_{s,p}}$    [m]    & 1.67e-08   & -   & \multicolumn{1}{c|}{-} & 1.67e-08  \\
			${\varepsilon_n}$    [-]       & 0.655   & -   & \multicolumn{1}{c|}{-} & 0.655  \\
			${\varepsilon_p}$    [-]    & 0.681   & -   & \multicolumn{1}{c|}{-} & 0.681  \\
			${A_{cell}}$    [${{\text{m}}^{\text{2}}}$]    & 2.125   & -   & \multicolumn{1}{c|}{-} & 2.125    \\
			${R_{l}}$    [${\Omega}$]      & 1.54e-03   & -   & \multicolumn{1}{c|}{-} & 1.54e-03 \\
			${D_{s,n}}$ [${{\text{m}}^{\text{2}}}{\text{/s}}$] & 1.28e-15   & 1.00e-10   & \multicolumn{1}{c|}{1.42e-15} & -        \\
			${D_{s,p}}$   [${{\text{m}}^{\text{2}}}{\text{/s}}$]            & 4.05e-18   & 5.45e-18   & \multicolumn{1}{c|}{2.74e-18} & -         \\
			${k_{n}}$  [${{\text{m}}^{{\text{2}}{\text{.5}}}}{\text{/(mo}}{{\text{l}}^{{\text{0}}{\text{.5}}}}{\text{s)}}$] & 2.02e-12   & 2.56e-12   & \multicolumn{1}{c|}{4.71e-12} & -  \\
			${k_{p}}$  [${{\text{m}}^{{\text{2}}{\text{.5}}}}{\text{/(mo}}{{\text{l}}^{{\text{0}}{\text{.5}}}}{\text{s)}}$]   & 9.50e-13   & 6.00e-13   & \multicolumn{1}{c|}{1.45e-12} & -         \\ \hline \hline
	\end{tabular}}
	\vspace{0em}
\end{table}

\begin{figure*}[!t]
	\centering
	\includegraphics[trim=0 10 0 0,clip,width = 2.05\columnwidth]{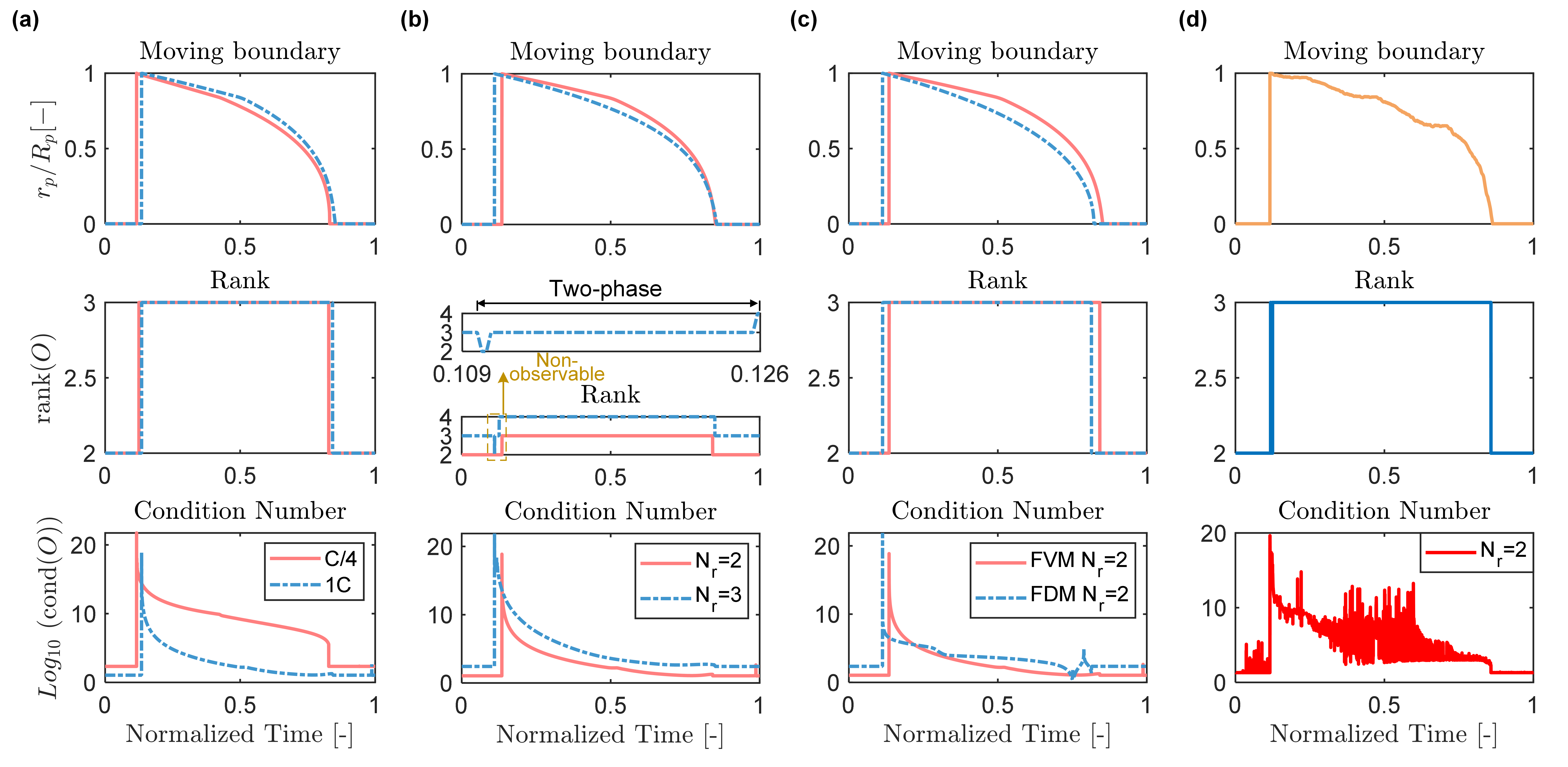}
	\caption{Observability analysis results containing phase transition dynamics, rank number and the common logarithm (i.e., $Log_{10}$) of condition number for (a) C/4 and 1C charge with $N_r=2$, (b) 1C charge with $N_r$=2 and $N_r$=3. \relx{The zoom plot shows the rank number between 10.9\% - 12.5\% SOC}, (c) FDM with $N_r=2$ and FVM with $N_r=2$ under 1C charge, and (d) UDDS current profile with $N_r=2$.}
	\label{fig:OB1}
	\vspace{-2em}
\end{figure*}

\section{Observability analysis} \label{Sec:Obs_analysis}
According to Section \ref{Sec:FVM_scheme} the resulting FVM-based discretized CSa-ESPM exhibits linearity in its states and nonlinearity in the output during one-phase conditions.
In the two-phase region, the state-space formulation Eq.\eqref{eq:tb5} becomes nonlinear in both state and output formualtion due to the dependency of $A_s$, $B_s$, and $G_s$ (Eq.\eqref{eq:tb9} - Eq.\eqref{eq:tb10}) on $r_p$.
Thus, in order to evaluate the feasibility of reconstructing the model states using the available input-output data, a nonlinear observability analysis is conducted. In this study, the rank condition is employed to assess the observability of the model in accordance with \cite{hermann1977nonlinear}.

\subsection{Non-linear observability analysis}
Let's consider a general non-linear model written as:
\begin{equation}
	\label{eq:OB1} 
\begin{array}{*{20}{c}}
	{\dot x = f\left( {x,u} \right)} \\ 
	{y = h\left( {x,u} \right)} 
\end{array}
\end{equation}
where $x \in {\mathbb{R}^n}$, $u \in {\mathbb{R}^m}$, $y \in \mathbb{R}$, $f:{\mathbb{R}^n} \times {\mathbb{R}^m} \to {\mathbb{R}^n}$, and $h:{\mathbb{R}^n} \times {\mathbb{R}^m} \to \mathbb{R}$.

\textbf{Definition 1 \cite{villaverde2018input}:}
Let $f:{\mathbb{R}^n} \times {\mathbb{R}^m} \to {\mathbb{R}^n}$ be a smooth vector field, $h:{\mathbb{R}^n} \times {\mathbb{R}^m} \to \mathbb{R}$ be a smooth scalar function, and $u \in {\mathbb{R}^m}$ be the system input. The J-order Extended Lie derivative of $h$ with respect to $f$ and $u$ is defined as:
\begin{equation}
	\label{eq:OB2} 
L_f^J\left( h \right) = \frac{{\partial L_f^{J - 1}\left( h \right)}}{{\partial x}} \cdot f + \sum\limits_{i = 0}^{J - 1} {\frac{{\partial L_f^{J - 1}\left( h \right)}}{{\partial {u^{\left( i \right)}}}}{u^{\left( {i + 1} \right)}}}
\end{equation}
Where $L_f^0\left( h \right) = h\left( {x,u} \right)$, and ${u^{\left( i \right)}}$ denotes the $i$-th order derivative of the input $u$.

\textbf{Theorem 1 \cite{hermann1977nonlinear}:}
The system \eqref{eq:OB1} is locally weakly observable at ($x_0$, $u_0$) if 
\begin{equation}
	\label{eq:OB3} 
rank\left( {{{\left. O \right|}_{\left( {{x_0},u_0} \right)}}} \right) = n
\end{equation}
Where $n$ is the number of states of the model and ${\left. O \right|_{\left( {{x_0},u_0} \right)}}$ is the observability matrix evaluated at ($x_0$, $u_0$) given by 
\begin{equation}
	\label{eq:OB4} 
{\left. O \right|_{\left( {{x_0},u_0} \right)}} = \frac{\partial }{{\partial x}}{\left[ {\begin{array}{*{20}{c}}
			{L_f^0\left( h \right)} \\ 
			{L_f^1\left( h \right)} \\ 
			\vdots  \\ 
			{L_f^{n - 1}\left( h \right)} 
	\end{array}} \right]_{({x_0},u_0)}}
\end{equation}

It is worth noting that the rank test \eqref{eq:OB4} determines whether the CSa-ESPM is weakly locally observable. 
However, it does not provide information about the accuracy or reliability of the estimates. 
To that end, the condition number of the observability matrix ($\kappa \left( {{{\left. O \right|}{\left( {{x_0},u_0} \right)}}} \right)$),  serves as a metric for evaluating the system observability, as outlined in \cite{allam2021linearized}. 
Specifically, $\kappa \left( {{{\left. O \right|}{\left( {{x_0},u_0} \right)}}} \right)$ is given by:
\begin{equation}
	\label{eq:OB7} 
\kappa \left( {{{\left. O \right|}_{\left( {{x_0},u_0} \right)}}} \right) = \left\| {{{\left. O \right|}_{\left( {{x_0},u_0} \right)}}^{ - 1}} \right\|\left\| {{{\left. O \right|}_{\left( {{x_0},u_0} \right)}}} \right\|
\end{equation}

The condition number of a matrix reflects its degree of ill-conditioning and proximity to singularity. 
Thus, a high condition number leads to state estimates with significant inaccuracies.

\subsection{CSa-ESPM observability analysis}
Prior research has established that constructing an observer intended to simultaneously estimate the concentrations of both electrodes from the available current and voltage data provides inaccurate outcomes attributed to limited observability. To address this challenge, one effective approach is to develop dedicated observers for each electrode individually \cite{9195005}. In this section, the non-linear observability of the positive electrode is studied. The states of the positive electrode in the one-phase and two-phase regions are described by Eq.\eqref{eq:tb1add} and Eq.\eqref{eq:tb6}, respectively.
Under the assumption of discretizing the particle solid-phase governing equations with 2 CVs, the extended Lie derivative for the positive electrode in the one-phase scenario \footnote{The same formulation holds also for the negative electrode \cite{allam2021linearized}} is given as follows:
\begin{equation}
	\label{eq:OB5} 
\begin{array}{l}
	L_f^0(h_{{\text{1P}}}) = {h_{{\text{1P}}}}\left( {{{{\mathbf{\bar c}}}_{s,p}},I} \right)= {U_p} + {\eta _p} \\ 
	L_f^1 (h_{{\text{1P}}}) = \frac{\partial  L_f^0(h_{{\text{1P}}})}{\partial {{{\mathbf{\bar c}}}_{s,p}}} \cdot ({f_\text{1P}}\left( {{{{\mathbf{\bar c}}}_{s,p}},I} \right)) + \frac{\partial  L_f^0(h_{{\text{1P}}})}{\partial I} \dot{I}\\
\end{array}
\end{equation}
where the positive OCP $U_p$ and the solid-phase overpotential ${\eta _p}$ are given by Eq. (8) - (9) in \cite{9991824}. ${f_\text{1P}}\left( {{{{\mathbf{\bar c}}}_{s,p}},I} \right)$ is the state-space equation for one-phase concentration given by Eq. (38) - (41) in \cite{Xu_2023}.
Note that the observability matrix, computed as in \eqref{eq:OB4}, is full rank within the one-phase region when the rank number is 2.
	In the two-phase scenario, alongside ${{\mathbf{\bar c}}_{s,p}}$, the model state also includes the presence of $r_p$. Writing the state vector in a compact form as ${{\mathbf{x}}_{2P}} = [{{\mathbf{\bar c}}_{s,p}}, r_p]$, the Lie derivative for the positive electrode in the two-phase scenario is then calculated as follows:
\begin{equation}
	\label{eq:OB6} 
\begin{array}{l}
	L_f^0({h_{{\text{2P}}}}) = {h_{{\text{2P}}}}\left( {{{\mathbf{x}}_{{\text{2P}}}},I} \right)= {U_p} + {\eta _p} \\ 
	L_f^1 ({h_{{\text{2P}}}}) = \frac{\partial  L_f^0({h_{{\text{2P}}}})}{\partial {{\mathbf{x}}_{{\text{2P}}}}} \cdot ({f_{{\text{2P}}}}\left( {{{\mathbf{x}}_{{\text{2P}}}},I} \right)) + \frac{\partial  L_f^0({h_{{\text{2P}}}})}{\partial I} \dot{I}\\
	L_f^2 ({h_{{\text{2P}}}}) = \frac{\partial  L_f^1({h_{{\text{2P}}}})}{\partial {{\mathbf{x}}_{{\text{2P}}}}} \cdot ({f_{{\text{2P}}}}\left( {{{\mathbf{x}}_{{\text{2P}}}},I} \right)) + \frac{\partial  L_f^0({h_{{\text{2P}}}})}{\partial I} \dot{I} + \frac{\partial  L_f^1({h_{{\text{2P}}}})}{\partial \dot{I}} \ddot{I} \\
\end{array}
\end{equation}
where ${f_{{\text{2P}}}}\left( {{{\mathbf{x}}_{{\text{2P}}}},I} \right)$ is the state-space equation for two-phase concentration given by Eq.\eqref{eq:tb5} - Eq.\eqref{eq:tb10} in Table \ref{Tab:POD}.
In this case, the rank of observability matrix must be 3 to ensure model observability.
Note that under constant current (CC) cycling conditions, both $\dot{I}$ and $\ddot{I}$ in \eqref{eq:OB5} and \eqref{eq:OB6} are zero. Additionally, to streamline the analysis in this study, we assume that the second and higher-order derivatives of the dynamic current input are negligible.

\begin{figure}[!t]
	\centering
	\includegraphics[trim=0 0 0 0,clip,width = 0.9\columnwidth]{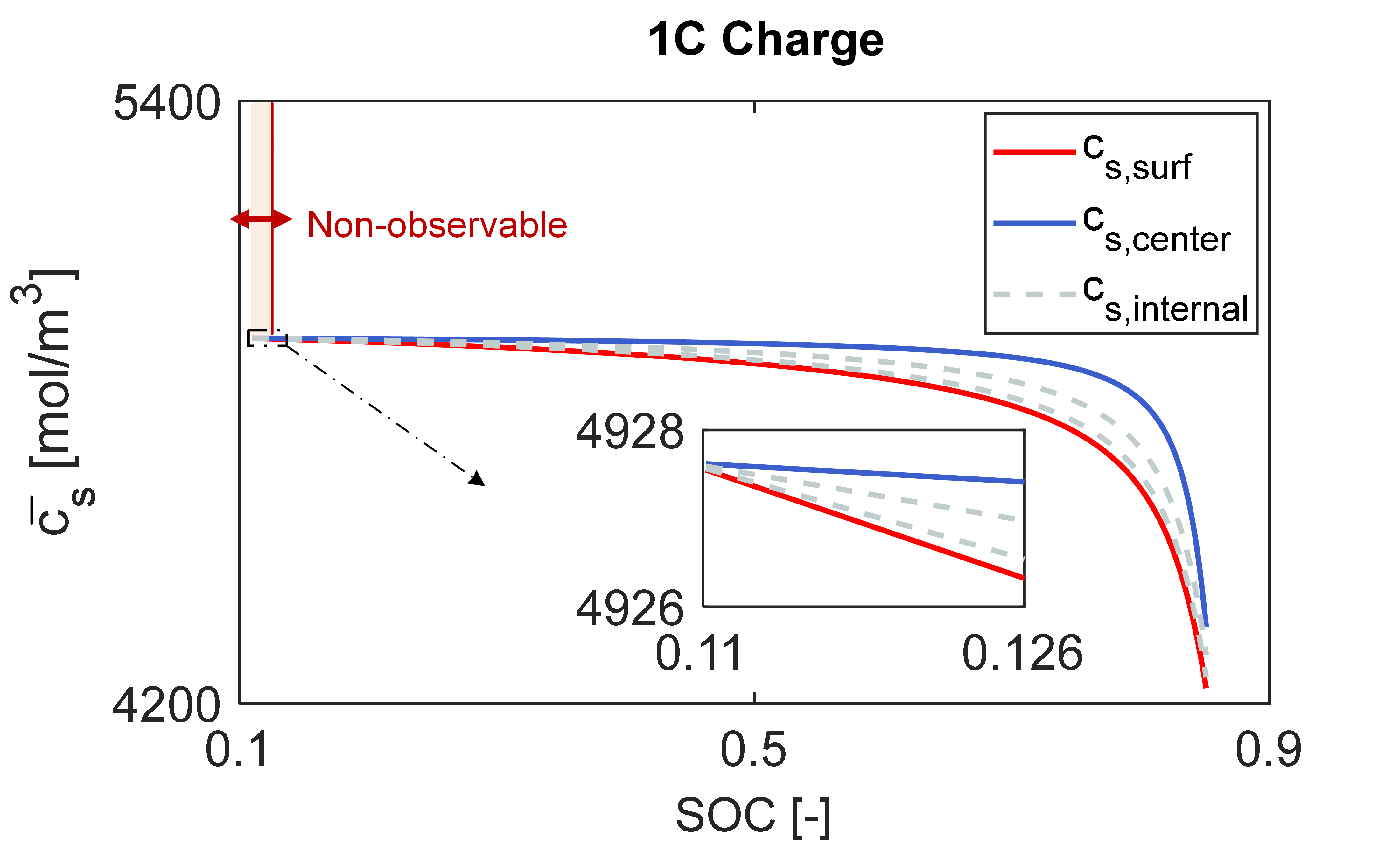}
	\caption{Positive electrode concentration distribution in two-phase region for FVM with $N_r=3$ during 1C charge}
	\label{fig:OB2}
	\vspace{-2em}
\end{figure}

An important aspect for developing electrode-based observer is to select the discretization grid points (i.e. CVs for FVM). 
FVM conserves mass by design, enabling the use of fewer control volumes in the discretization of CSa-ESPM.
In Figure \ref{fig:OB1}(a), the observability results of a CSa-ESPM with 2 CVs under constant charge currents of C/4 and 1C are depicted. The observability matrix of the positive electrode remains at full rank, and the condition number increases as the magnitude of current decreases.
Furthermore, Figure \ref{fig:OB1}(b) examines the nonlinear observability of the positive electrode during a 1C charge for $N_r=2$ and $N_r=3$. In the $N_r=3$ scenario, the observability matrix experiences a loss of rank (rank of 3 instead of 4 in the two-phase region), particularly at the beginning of the two-phase region (SOC between 10.9\% to 12.5\%), rendering the positive electrode unobservable.
The observed loss of rank is attributed to the minimal concentration difference, and will be discussed later. Additionally, the condition number increases as the number of CVs rises. 
Next, a comparison of positive electrode nonlinear observability between FDM and FVM versions of the CSa-ESPM  is presented during a 1C charge  in Figure \ref{fig:OB1}(c). Notably, FVM exhibits a lower overall condition number compared to FDM, suggesting that the FVM-based CSa-ESPM observer offers enhanced accuracy.
Finally, Figure \ref{fig:OB1}(d) illustrates the nonlinear observability of the CSa-ESPM positive electrode under the UDDS dynamic current input profile. Compared to CC input, the condition number plot exhibits more fluctuations, suggesting that observer accuracy is likely to decrease under dynamic current conditions.
In Fig. \ref{fig:OB1}(b) it was observed that the CSa-ESPM positive electrode becomes unobservable at the beginning of the two-phase region when $N_r=3$. The explanation for this finding can be understood by analyzing the positive electrode concentration distribution in the two-phase region, as illustrated in Figure \ref{fig:OB2} for 1C CC charging cycle.
Under this condition, when the LFP positive electrode initially enters the two-phase region, the shell thickness is minimal, resulting in a negligible concentration difference between the particle surface and center. Consequently, such a small disparity cannot be precisely reconstructed based solely on the measured voltage and current. This trend is expected to exacerbate with an increase in the number of CVs,  as it improves the level of detail within the model.

\section{Conclusion}\label{sec:Conc}
In this paper, the FVM is applied to spatially discretize the CSa-ESPM built for LFP batteries. The resulting state-space model greatly reduces the solid-phase state variables to less than 6 ($N_r$=4 for ${{\mathbf{\bar c}}_{s,p}}$ plus $r_p$) while guaranteeing mass conservation. As shown from experimental validation results, the reduced-order CSa-ESPM matches well with both cell voltage and electrode SOC. Moreover, CSa-ESPM observability is analyzed and quantified by running the rank test and using condition numbers. The proposed FVM-based CSa-ESPM together with the nonlinear observability analysis is a first step for the development of control-oriented models for electrode-based observers in BMS applications.

\vspace{-0.5em}
\bibliographystyle{IEEETran}

\end{document}